\documentclass{article}
\usepackage{graphicx}
\usepackage{amsmath}
\usepackage{amssymb}
\usepackage{color}

\def\cow{\color{white}}

\begin{document}

\title{Patch cosmology and noncommutative braneworlds\footnote{Talk presented at DESY Theory Workshop 2004, Hamburg, 27 Sep.- 1 Oct. 2004, {\it Particle Cosmology}.}}

\author{Gianluca Calcagni\footnote{calcagni@fis.unipr.it}\\
Dipartimento di Fisica, Universit\`a di Parma\\
INFN -- Gruppo collegato di Parma \\ 
Parco Area delle Scienze 7/A, I-43100 Parma, Italy}

\maketitle

\begin{abstract}
We review extra-dimensional and 4D cosmological scenarios through the effective Friedmann evolution on a brane. Some features involving noncommutative geometry and scalar/tachyon slow-roll inflation are considered.
\end{abstract}


\section{Introduction}

Motivated by recent developments in string, superstring and M theory, several models for a multidimensional target spacetime have been proposed. Among them, particular attention has been devoted to brane-world scenarios, according to which the visible universe is a (3+1)-dimensional variety (a 3-brane) embedded in a bulk with some either non-compact or compactified extra dimensions. Typically, the background metric on the brane is assumed to be the Friedmann-Robertson-Walker (FRW) metric and the Einstein equations are modified in accordance with the gravity model describing spacetime. Their projection on the brane results in the basic FRW equations for the cosmological evolution.  
For an introduction to the subject and some lists of references, see \cite{bw}.

In this paper we review how to look for cosmic signatures of high-energy, higher-derivative gravity models. In particular, the construction of a nontrivial set of consistency equations permits us to compare theoretical predictions with the perturbation spectra of the cosmic microwave background (CMB). It turns out that CMB experiments of this and next generation might be able to discriminate between the standard four-dimensional lore and the braneworld paradigm. The introduction of a fundamental energy scale reinforces this result. In Sec. \ref{setup} we introduce the basic ingredients of the patch, slow-roll and noncommutative formalisms, taking as examples the five-dimensional Randall-Sundrum (RS) scenarios and their Gauss-Bonnet (GB) generalization. In Sec. \ref{pert} we outline some results on cosmological perturbations on a noncommutative brane and find their observational consequences. Conclusions are in Sec. \ref{concl}


\section{Setup} \label{setup}

One of the first problems one has to deal with when constructing braneworld models is how to stabilize the extra dimension. This can be achieved in a number of ways; in the RS example, Goldberger and Wise have provided a mechanism according to which a 5D massive scalar is put into the bulk with a potential of the same order of the brane tension $\lambda$ \cite{GW}. If the energy density $\rho$ on the brane is smaller than the characteristic energy of the scalar potential, $\rho/V \sim \rho/\lambda \ll 1$, then the radion is stabilized and one gets the standard Friedmann equation $H^2 \propto \rho$ on the brane. On the contrary, if the brane energy density is comparable with the stabilization potential, $\rho/\lambda \gtrsim 1$, the bulk backreacts because it feels the presence of the brane matter, the minimum of the potential is shifted and the well-known quadratic corrections to the Friedmann equation arise:
\begin{equation}
H^2 = \frac{\kappa_4^2}{6 \lambda} \rho (2 \lambda+\rho)\,,	
\end{equation}
where $H$ is the effective Hubble rate experienced by an observer on the brane and $\kappa_4$ is the 4D gravitational coupling. 

The RS model can be viewed as an intermediate scenario between a ``pure Gauss-Bonnet'' high-energy regime, $H^2 \propto \rho^{2/3}$, and the standard 4D (low-energy) evolution, $H^2 \propto \rho$. Here we shall consider nonstandard cosmological evolutions on the brane and extend the RS and GB discussion to arbitrary scenarios we dubbed ``patch cosmologies'' \cite{cal3}, with
\begin{equation} \label{FRW}
H^2=\beta_q^2 \rho^q\,,
\end{equation}
where $\beta_q$ is a constant and the exponent $q$ is equal to 1 in the pure 4D (radion-stabilized) regime, $q=2$ in the high-energy limit of the RS braneworld and $q=2/3$ in the high-energy limit of the GB scenario. In order to simplify the framework, we make the following assumptions:
\begin{enumerate}
\item There is a confinement mechanism such that matter lives on the brane only, while gravitons are free to propagate in the bulk. This is guaranteed as long as $\rho<M_5^4$.
\item The contribution of the Weyl tensor is neglected.
\item The contribution of the anisotropic stress is neglected.
\item We concentrate on the large-scale limit of the cosmological perturbations.
\end{enumerate}
Assumption 2 closes the system of equations on the brane and sets aside the nonlocal contributions from the bulk,\footnote{To neglect the projected Weyl tensor implies that there is no brane-bulk exchange. The converse is not true: given a standard continuity equation on the brane, the Friedmann equation still can get an extra dark-radiation term (e.g. \cite{KKTTZ}).} while 
assumptions 3 and 4 reduce the number of degrees of freedom of gauge invariant scalar perturbations. 

This list might seem too restrictive and to spoil almost all the interesting features of the model. However, assumptions 4 and 2 nicely fit in the inflationary regime, since the long wavelength region of the spectrum, corresponding to the Sachs-Wolfe plateau, encodes the main physics of the inflationary era. Moreover, the dark radiation term, which is the simplest contribution of the Weyl tensor, scales as $a^{-4}$ and is exponentially damped during the accelerated expansion. Finally, bulk physics mainly affects the small-scale/late-time cosmological structure and can be consistently neglected during inflation. This is a highly nontrivial result which has been confirmed with several methods both analytically and numerically \cite{bu}.

Imposing a perfect fluid on the brane with equation of state $p=w\rho$, the continuity equation governing the cosmological dynamics is the same as in four dimensions, thanks to assumption 2:
\begin{equation}
\dot{\rho}+3H (\rho+p)=0\,.
\end{equation}
There are two candidates for the role of inflation. The first one is an ordinary scalar field $\phi$ with energy density and pressure
\begin{equation}
\rho=\dot{\phi}^2/2+V(\phi)=p+2V(\phi)\,.
\end{equation}
The second one is a Dirac-Born-Infeld (DBI) tachyon $T$ such that
\begin{eqnarray} \label{tac}
\rho &=& V(T)/c_S=-V(T)^2/p\,,\\
c_S &\equiv& \sqrt{1-\dot{T}^2}\,.
\end{eqnarray}
From a string-theoretical point of view, the evolution of the tachyon proceeds up to the condensation $\dot{T}\to 1$ into the closed string vacuum, where no signal of open excitations propagates (Carrollian limit). Also, near the minimum the strong coupling regime emerges, $g_s=O(1)$, and the perturbative description implicit in the DBI action may fail down. However, from a cosmological perspective Eq. (\ref{tac}) is a toy model and, like in standard inflation, an additional reheating mechanism around the condensation is required for gracefully exiting the inflationary period. Here we will not consider this and other (indeed solvable) problems concerning the tachyon and just implement the DBI action in the cosmological dynamics as an alternative model of inflation.

\subsection{Slow-roll parameters}

Let $\psi$ denote the inflaton field irrespectively of its action. Expressions involving $\psi$ will be valid for both the ordinary scalar and the tachyon. The first-order slow-roll (SR) parameters are defined as
\begin{equation} \label{1SR}
\epsilon \equiv -\frac{\dot{H}}{H^2}\,,\qquad \eta \equiv -\frac{\ddot{\psi}}{H\dot{\psi}}\,,
\end{equation}
together with their evolution equations with respect to synchronous time
\begin{equation}
\dot{\epsilon} = H\epsilon \left[\left(2-\widetilde{\theta}\right)\,\epsilon-2\eta\right]\,,\qquad
\dot{\eta}     =  H\left(\epsilon\eta-\xi^2\right)\,,
\end{equation}
where $\xi^2 \equiv (\ddot{\psi}/\dot{\psi})^\cdot/H^2$ is a second-order parameter, in the sense that it appears only in expressions which are $O(\epsilon^2,\eta^2,\epsilon\eta)$. Here $\widetilde{\theta}=2$ for the tachyon field and $\widetilde{\theta}=\theta \equiv 2(1-q^{-1})$ for the ordinary scalar field (4D: $\theta=0$; RS: $\theta=1$; GB: $\theta=-1$). Note that each time derivative of the SR parameters increases the order of the SR expressions by one.

The first SR parameter is actually the time derivative of the Hubble radius $R_H\equiv H^{-1}$. Because of its purely geometrical content, it cannot be implemented in these SR towers recursively. By definition, there is inflation when $\epsilon < 1$:
\begin{equation} \label{infl}
\frac{\ddot{a}}{a} = H^2 (1-\epsilon)\,.
\end{equation}
Under the \emph{slow-roll approximation}, if the potential term dominates over the kinetic term, then the inflaton slowly rolls down its potential, $\epsilon,\eta \ll 1$, and the perfect fluid mimics that of a cosmological constant, $p \approx -\rho$. Deviations from the de Sitter behaviour generate large-scale perturbations which explain the anisotropies in the CMB. 

One can construct infinite towers of SR parameters encoding the full dynamics of the inflationary model. For instance, Eq. (\ref{1SR}) provides the first entries of the ``Hubble" SR tower; another, sometimes more convenient tower is the potential tower defined in \cite{cal3} and related to the Hubble tower by approximated relations.


\subsection{The noncommutative brane}

Until now we have considered a commutative background throughout the whole spacetime. We can make a step further and phenomenologically assume to have a 3-brane in which the stringy spacetime uncertainty relation (SSUR)
\begin{equation} \label{SSURph}
\Delta \tau \Delta x \geq l_s^2\,,
\end{equation}
holds for all the braneworld coordinates $\{x^\nu\}$, $\nu = 0,1\,2\,3$, while the extra dimension $y$ along the bulk remains decoupled from the associated *-algebra. Here, $\tau \equiv \int a\,dt$ ($\approx a/H$ in the SR regime), $x$ is a comoving spatial coordinate on the brane and $l_s \equiv M_s^{-1}$ is the fundamental string scale. 

Noncommutative braneworld inflation arises when we impose a realization of the 
*-algebra on the brane coordinates. In order to diagonalize the noncommutative algebra and induce a pure 4D SSUR on the brane, one might fix the expectation values of the background fields of the fundamental theory such that the extra direction commutes, $[y,x^\nu]=0$. An algebra preserving the maximal symmetry of the FRW universe is $[\tau,x]=il_s^2$ \cite{BH}, by which normal products in the action of the perturbation modes are replaced by *-products like
\begin{equation}\label{*prod}
(f*g)(x,\tau)=e^{-(il_s^2/2)(\partial_x\partial_{\tau'}-\partial_{\tau}\partial_{x'})}f(x,\tau)g(x',\tau')\big|_{\text{\tiny \begin{tabular}{l} $x'=x$ \\ $\tau'=\tau$\end{tabular}}}\,.
\end{equation}


\section{Cosmological perturbations: theory and observations} \label{pert}

\subsection{Commutative case}

Quantum fluctuations of the scalar field governing the accelerated era are inflated to cosmological scales because of the superluminal expansion. They constitute the seeds of both the small anisotropies observed in the microwave sky and the large-scale nonlinear structures around which gravitating matter organizes itself. For an introduction of the subject in the general relativistic case, see \cite{MFB}. The standard procedure to adopt in order to compute the perturbation spetrum is: (a) Write the linearly perturbed metric in terms of gauge-invariant scalar quantities. (b) Compute the effective action of the scalar field fluctuation and the associated equation of motion. (c) Write the perturbation amplitude as a function of an exact solution of the equation of motion with constant SR parameters. (d) Perturb this solution with small variations of the parameters.

In scenarios with an extra dimension the full computation is very nontrivial due to either the extra degrees of freedom in the 5D metric and the complicated geometrical background on which to solve the Einstein equations coupled with the junction conditions on the brane. However, as explained above things become simpler when going to the large-scale limit. In this case, several arguments show that the resulting spectra are, to lowest SR order,
\begin{eqnarray}
A       &=& \frac{k}{5\pi z}\,,\\
z(\phi) &=& \frac{a\dot{\phi}}{H}\,,\\
z(T)    &=& \frac{a\dot{T}}{c_S\beta_q^{1/q} H^{\theta/2}}\,,\\
z(h)    &=& \frac{\sqrt{2}a}{\kappa_4 F_q}\,,\\
F^2_q &\equiv& \frac{3q\beta_q^{2-\theta}H^\theta}{\zeta_q\kappa_4^2}\,,
\end{eqnarray}
where $A(h)=A_t$ is the tensor spectrum of the gravitational sector and $\zeta_q$ is a numerical constant which depends on the concrete gravity model one is considering: it is $\zeta_1=1=\zeta_{2/3}$ for the 4D and GB cases and $\zeta_2=2/3$ for RS \cite{zeta,cal5}. 

To lowest order, the scalar and tensor spectral indices are first order in the SR parameter, $n_t \equiv d \ln A_t^2/d \ln k \sim O(\epsilon) \sim n_s-1 \equiv d\ln A_s^2/d \ln k$, while their running $\alpha_{s,t} \equiv d n_{s,t}/d \ln k$ is second order. Here $k$ is the comoving wave number of the perturbation and the subscripts $s$ and $t$ refer to scalar and tensor perturbations, respectively. In the case of exact scale invariance, $n_s=1$ and $n_t=0$. The tensor-to-scalar ratio is
\begin{equation}
r \equiv A_t^2/A_s^2=\epsilon/\zeta_q+O(\epsilon^2)\,.
\end{equation}
Combining the SR expressions of the observables, one gets the consistency equations
\begin{subequations}\label{ce}
\begin{eqnarray}
n_t            &=& -(2+\theta)\zeta_q r+O(\epsilon^2)\,, \label{1ce}\\
\alpha_t &=& (2+\theta)\zeta_qr[(2+\theta)\zeta_qr+(n_s-1)]\,,\\
\alpha_s(\phi) &\approx& \zeta_q r [4(3+\theta)\zeta_qr+5(n_s-1)]\,,\\
\alpha_s(T)    &\approx& (3+\theta)\zeta_q r [(2+\theta)\zeta_qr+(n_s-1)]\,.
\end{eqnarray}
\end{subequations}
The key point is that the set of consistency relations is not degenerate when considering different patches $\theta$ and $\theta'$. The only known (accidental) degeneracy is for Eq. (\ref{1ce}) in the RS and 4D case, where $n_t=-2r$ at first SR order. However, the second-order version of this equation together with the expressions for the runnings definitely break the degeneracy. This implies that, at least in principle, braneworld scenarios can be discriminated between each other.

To quantify the effect of the extra dimension, we can use the recent CMB data coming from WMAP \cite{wmap}. With the upper bound $r < 0.06$ for the tensor-to-scalar ratio and the best-fit value $n_s \approx 0.95$ for the scalar spectral index, the relative scalar running in two different patches is
\begin{equation} \label{deltalp}
\alpha_s^{(\theta,\psi)}-\alpha_s^{(\theta',\psi')} \sim O(10^{-2})\,, 
\end{equation}
which is close to the WMAP estimate of the experimental error. This estimate will be highly improved by either the updated WMAP data set and near-future experiments, including the European Planck satellite, for which the forecast precision should be ameliorated by one order of magnitude, $\Delta\alpha_s \sim O(10^{-3})$. 

\subsection{Noncommutative case}

Let us introduce the noncommutative parameter $\mu \equiv (H/M_s)^4$; the noncommutative algebra 
induces a cutoff $k_0(\mu)$ roughly dividing the space of comoving wave numbers into two regions, one encoding the UV, small-scale perturbations generated inside the Hubble horizon ($H \ll M_s$) and the other describing the IR, large-scale perturbations created outside the horizon ($H \gg M_s$). By definition, they correspond to 
the quasi-commutative and strongly noncommutative regime, respectively. Since the *-product (\ref{*prod}) does not involve homogeneous quantities, the exact solutions of the commutative equations of motion do hold for the noncommutative case, too, and it turns out that one can factorize the nonlocal effects in the amplitudes and write the latter ones as
\begin{equation}
\label{Anoncom}
A(\mu,\,H,\,\psi) = A^{(c)} (H,\,\psi)\,\Sigma (\mu)\,,
\end{equation}
where the subscript $(c)$ denotes quantities in the commutative limit [$A^{(c)}=A(\Sigma\!\!=\!\!1)$] and $\Sigma(\mu)$ is a function encoding leading-SR-order noncommutative effects. With amplitudes of this form, $r=r^{(c)}$.

Equation (\ref{Anoncom}) is evaluated at horizon crossing in the UV limit and at the time when the perturbation with comoving wavenumber $k$ is generated in the IR limit. To lowest SR order \cite{cal4},
\begin{eqnarray}
\frac{d \ln \Sigma^2}{d \ln k} = \sigma \epsilon\,,\\
n = n^{(c)}+\sigma\epsilon\,,
\end{eqnarray}
where $\sigma = \sigma(\mu)$ is a function of $\mu$ such that $\dot{\sigma}=O(\epsilon)$. The standard commutative spacetime corresponds to $\sigma=0$.

In the UV limit ($\mu \ll 1$), the noncommutative part of the amplitudes can be written as
\begin{eqnarray}
\Sigma^2     &\approx& 1-b\mu\,,\\
\sigma       &\approx& 4b\mu\,,
\end{eqnarray}
where $b$ is a numerical coefficient shown in table \ref{tab1} for the models considered in \cite{cal4} (see this paper for details). In the first class (``1''), the FRW 2-sphere is factored out in the measure of the effective 4D perturbation action $z_k$, which is given by the product of the commutative measure $z$ and a correction factor from the $(1+1)$-dimensional noncommutative action. In the class 2 choice, the scale factor in the measure is everywhere 
substituted by an effective scale $a_{\text{\tiny eff}}$ whose time dependence is smeared out by the nonlocal physics; since $z \propto a$, then $z_k=z a_{\text{\tiny eff}}/a$. Inequivalent prescriptions on the ordering of the *-product in the perturbation action further split these two classes, one corresponding to the Brandenberger-Ho model (BH) and the other one described in \cite{cal4}, but in the IR limit they give almost the same predictions. In this regime, the function $\sigma$ is asymptotically constant, $\lim_{\mu \to \infty}\sigma(\mu)$=const; this nontrivial feature will permit us to perform an analysis of the cosmological data without enlarging the parameter space.
\begin{table}[ht]
\caption{\label{tab1} Noncommutative perturbation amplitudes in the UV and IR limits.}\begin{center}
\begin{tabular}{l|cc||l|cc}
        &   UV    & & \multicolumn{2}{c}{IR}\\
        &   $b$   & &   $\Sigma^2$   & $\sigma$ \\ \hline
BH1     &    4    & & $\mu^{-3/2}/2$ &     6   \\
New1    &  3/2    & & $\mu^{-3/2}$   &     6   \\
BH2     &    1    & &  $\mu^{-1/2}$  &     2   \\
New2    &  1/2    & &  $\mu^{-1/2}$  &     2   \\
\end{tabular}\end{center}\end{table}

The commutative consistency equations (\ref{ce}) can be generalized in a straightforward way. An easy calculation shows that all degeneracies are removed, including the RS-4D one \cite{cal5}. The relation between the tensor index and the amplitude ratio now reads
\begin{equation}\label{cenc}
n_t = [\sigma-(2+\theta)] \zeta_q r\,,
\end{equation}
with IR values displayed in Table \ref{tab2}. Note that the positive contribution of $\sigma$ allows blue-tilted scalar and tensor spectra. Moreover, the introduction of a fundamental length scale allows a greater range for the relative running (\ref{deltalp}), up to $\Delta\alpha_s \sim O(10^{-1})$ in some cases \cite{cal4}.
\begin{table}[ht]
\caption{\label{tab2} The consistency relation (\ref{cenc}) in the IR limit.}\begin{center}
\begin{tabular}{c|ccc}
 Consistency                        &\multicolumn{3}{c}{$n_t/r$}\\
 relation                           &       GB     &           RS    &   4D   \\ \hline
Commutative UV ($\sigma=0$) &      $-1$      &          $-2$     &   $-2$   \\
Class 1 IR ($\sigma=6$)     &       {\cow +}5      &           {\cow +}2     &   {\cow +}4   \\ 
Class 2 IR ($\sigma=2$)     &       {\cow +}1      &  $-2/3$   &   {\cow +}0   \\
\end{tabular}\end{center}\end{table}

\begin{figure}\begin{center}
\includegraphics[height=4in,width=4in]{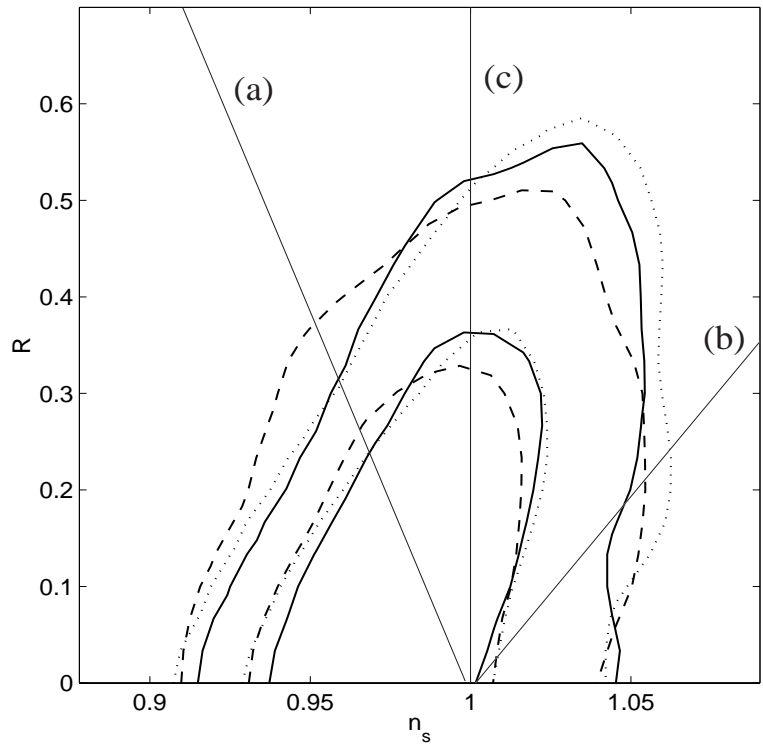}\end{center}
\caption{\label{fig2}
The $1\sigma$ and $2\sigma$ observational contour bounds for the 4D case, where $R=16r$. Each contour curve corresponds to $\sigma=0$ (solid), $\sigma=6$ (dashed) and $\sigma=2$ (dotted). We also show the border of large-field (left) and hybrid (right) inflationary models \cite{CT}.}
\end{figure}

In \cite{CT} the Cosmological Monte Carlo (CosmoMC) code together with CAMB were run \cite{cmc} using the data set of WMAP coupled to that of other experiments (2dF and SDSS, plus CBI, VSA and ACBAR for small scales \cite{data}). For each noncommutative patch ($\theta$ and $\sigma$ fixed) we have carried out a likelihood analysis valid for both the ordinary scalar and the tachyon. As an example, Fig. \ref{fig2} shows the likelihood contours in the general-relativistic case and the effect of noncommutativity. In general, blue-tilted spectra are allowed and the constraints on the inflaton potential deeply change, see \cite{CT}.

\begin{figure}\begin{center}
\includegraphics[height=4in,width=4in] {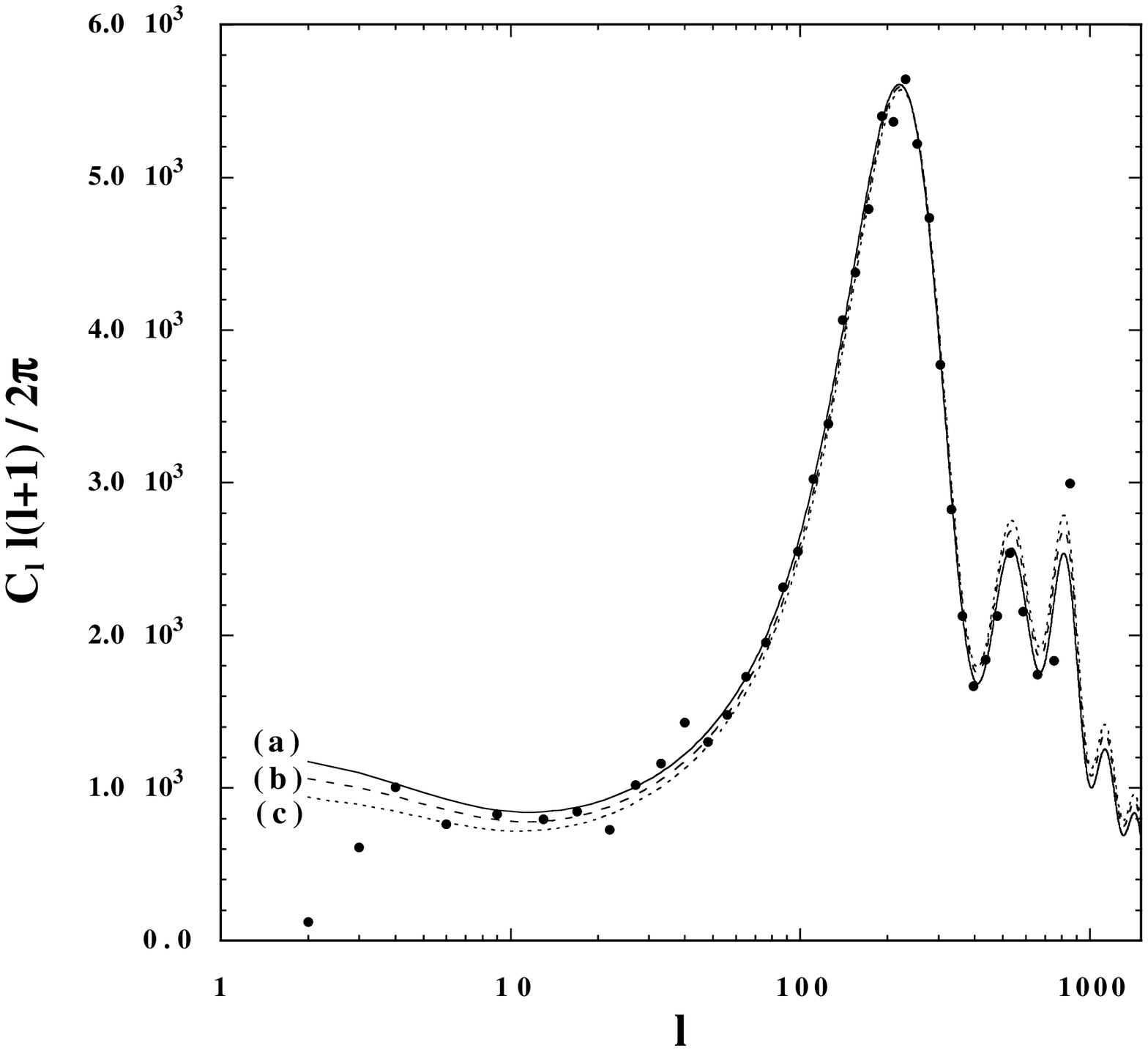}\end{center}
\caption{\label{fig11} The CMB angular power spectrum showing the effects of
suppression of power at low multipoles. 
Curve~(a) is the 4D commutative normal scalar model
with $(n_s, R)=(0.967, 0.132)$ ($V=\phi^2$).
Curves~(b) and (c) are the 4D-1 noncommutative scenario with
$(n_s, R)=(1.018, 0.144)$ ($V=\phi^2$) and 
$(n_s, R)=(1.049, 0.263)$ ($V=\phi^4$), respectively \cite{CT}.}
\end{figure}

The last feature we want to stress is a mild suppression of the low CMB multipoles with respect to the commutative case, see Fig. \ref{fig11}. Although noncommutativity does not fully explain the data (nor can other exotic theories with even much bluer spectra), it is interesting to see how the smearing of spacetime at microscopic level is amplified to the large cosmological scales by inflation.


\section{Conclusions} \label{concl}

In this paper we have summarized some results on braneworld inflation and their observable consequences. We have not presented a full 5D calculation but we expect that bulk physics would not dramatically improve large-scale results \cite{bu}. The study of the microwave background could give the first clues of a wider spacetime in the next years or even months.
 
In addition to the brane conjecture, one may insert other exotic ingredients, borrowed from string and M theory, that may give rise to characteristic predictions, although at the price of increasing the number and complexity of concurring models. For instance, the introduction of a noncommutative scale can generate a blue-tilted spectrum and explain, at least partially, the low-multipole suppression of the CMB spectrum detected by WMAP; also, it modifies the observationally allowed regions in the parameter space.

It would be interesting to find new cosmological scenarios with $\theta\neq 0,\pm 1$ and exploit the compact formalism provided by the patch formulation of the cosmological dynamics. Certainly there could be a lot of work for M/string theorists in this direction.

A final important question is in order: Will the CMB be the smoking gun of extra dimensions or noncommutative scenarios? In the context of the patch formalism the answer, unfortunately, is no. Some general relativistic models may predict a set of values for the observables $\{n_t,n_s,r,\alpha_s,\dots\}$  close to that of a braneworld within the experimental sensitivity. Even noncommutativity may not escape this ``cosmic degeneracy" since, for example, a blue-tilted spectrum can be achieved by the 4D hybrid inflation. So we can talk about clues but not proofs about high-energy cosmologies when examining the experimental data. The subject has to be further explored in a more precise way than that provided here in order to find out more compelling and sophisticated predictions, extending the discussion also to the small-scale region of the spectrum.


\section*{Acknowledgments}
I would like to thank the organizers of DESY 2004 for their kind hospitality. I also acknowledge Shinji Tsujikawa for the figures appeared here and in \cite{CT}.


\end{document}